\documentclass{article}

\pdfoutput=1

\usepackage{spconf}

\usepackage{graphicx}
\usepackage{amsmath,amssymb}
\usepackage{color}

\usepackage{enumitem}
\usepackage{diagbox}
\usepackage{booktabs}
\usepackage{multirow}
\usepackage{colortbl}
\usepackage{xspace}
\usepackage{xcolor}
\usepackage{soul}
\usepackage{hyperref}
\usepackage{subcaption}

\usepackage{microtype}
\usepackage[capitalise]{cleveref}

\DeclareMathOperator*{\argmax}{arg\,max}

\title{
Mask The Bias: Improving Domain-Adaptive Generalization of CTC-based ASR with Internal Language Model Estimation
}
\name{
\begin{tabular}{c}
Nilaksh Das,
Monica Sunkara,
Sravan Bodapati,
\\
Jinglun Cai,
Devang Kulshreshtha,
Jeff Farris,
Katrin Kirchhoff
\end{tabular}
\vspace{-0.8em}
}
\address{AWS AI Labs, USA}

\begin{document}
\ninept

\maketitle

\begin{abstract}
End-to-end ASR models
trained on large amount of data 
tend to be implicitly biased 
towards language semantics of the training data.
Internal language model estimation (ILME)
has been proposed
to mitigate this bias
for autoregressive models
such as attention-based encoder-decoder and RNN-T.
Typically, ILME is performed by modularizing
the acoustic and language components
of the model architecture,
and eliminating the acoustic input
to perform log-linear interpolation
with the text-only posterior.
However,
for CTC-based ASR,
it is not as straightforward 
to decouple the model
into such acoustic and language components,
as CTC log-posteriors are computed in
a non-autoregressive manner.
In this work,
we propose a novel ILME technique
for CTC-based ASR models.
Our method iteratively masks 
the audio timesteps
to estimate a pseudo log-likelihood
of the internal LM
by accumulating
log-posteriors
for only the masked timesteps.
Extensive evaluation
across multiple out-of-domain datasets
reveals that the proposed approach
improves 
WER by up to 9.8\%
and OOV F1-score
by up to 24.6\% relative to Shallow Fusion,
when only text data from target domain is available.
In the case of zero-shot domain adaptation, 
with no access to any target domain data,
we demonstrate that removing the source domain bias 
with ILME can still outperform Shallow Fusion 
to improve WER by up to 9.3\% relative.
\end{abstract}


\section{Introduction} \label{sec:intro}
\vspace{-0.5em}

Automatic speech recognition (ASR)
has now become tightly integrated
with daily life
through commonly used real-world applications
such as 
digital assistants, 
news transcription 
and AI-based interactive voice response telephony.
Often in such practical scenarios,
ASR models
that are trained on some source domain training data
are deployed for other target domains
in which target domain training data, 
especially paired audio-text data, 
may be severely limited.
External language model (LM) integration techniques
have been proposed
to overcome such cases of domain mismatch,
wherein text-only data from the target domain is 
leveraged~\cite{toshniwal2018comparison,shan2019component,shenoy2021adapting,shenoy2021asr,liu2021domain,dingliwal2022domain}.
Complementary to adding an external target domain LM
to capture target domain semantics,
density ratio estimation techniques have been proposed
that aim to minimize any bias from the source domain
by utilizing an 
additional source domain LM~\cite{mcdermott2019density,choudhury2022likelihood,zhou2022language}.
Building further on top of this,
a growing body of recent works 
have also looked into internal LM estimation (ILME),
aimed at reducing any internal model bias
from the source domain 
by estimating implicit language semantics
learned internally by the model itself~\cite{meng2021internal,zeineldeen2021investigating,liu2022internal}.

ILME proposes to reduce source domain model bias
by modularizing the acoustic model (AM) and language model (LM) components 
of the model architecture, and
eliminating the acoustic input altogether.
Then, a log-linear interpolation with
the overall log-likelihood scores is performed
to subtract away the text-only log-posteriors~\cite{meng2021internal}.
Such an ILME approach is
fundamentally limited to 
autoregressive models
such as attention-based encoder-decoder (AED)
and recurrent neural network transducer (RNN-T),
as it is relatively easy to
decouple and approximate such architectures 
into AM and LM components.

Recently, models based on 
connectionist temporal classification (CTC) loss 
have demonstrated state-of-the-art 
ASR performance with high inference speed 
and low memory footprint~\cite{lee2021intermediate,fan2021cass,zhang2021benchmarking,dingliwal2022personalization}. 
%
%
However, the proposed ILME techniques
cannot be formulated readily
for CTC-based ASR models.
Given CTC's non-autoregressive architecture,
decomposing a CTC model into AM and LM components
is ambiguous.
In this work, we propose a novel ILME technique
for CTC models.
Our approach is to iteratively mask
the input audio timesteps
in order to estimate a pseudo log-likelihood
of the internal LM (ILM)
by accumulating
the log-posteriors
for only the masked timesteps
across all iterations.
This 
brings forth the ILM bias
by eliminating the acoustic input
in an iterative non-autoregressive fashion,
thus bypassing the need to
decouple the model
into an AM and LM.

\begin{table}[t!]
\footnotesize
\captionsetup{font=footnotesize}
\centering
\caption{Example of a CTC model's internal bias. With words masked from input audio, the model still attempts to decode a coherent output.}
\vspace{-1em}
\begin{tabular}{@{}l@{}l@{}}
\toprule
REF & he is a member of {\color{red} the} republican main {\color{red} street} partnership \\
HYP: greedy & he is a member of the republican main street partnership \\
HYP: mask ``the" & he is a member of {\color{blue} public and} main street partnership \\
HYP: mask ``street" \ & he is a member of the republican {\color{blue} mainstream} partnership \\
\bottomrule
\end{tabular}
\label{tab:ilm-example}
\vspace{-2em}
\end{table}

\noindent
Through this work, we make the following contributions:
\begin{itemize}[label={$\bullet$},leftmargin=*,topsep=1pt,itemsep=0pt]
\item \textbf{Novel ILME technique for CTC-based ASR.}
While ILME has mainly been proposed 
in the context of autoregressive models
like AED and RNN-T,
we present
a novel masking-based ILME technique
for non-autoregressive CTC decoding.
We experiment with multiple masking methods
including masking the audio at word boundaries. 
We find that simply dividing the audio 
into equal lengths
and masking each partition yields best results.
%
%
\item \textbf{Extensive evaluation across multiple out-of-domain datasets.}
We evaluate the proposed ILME approach 
across 4 out-of-domain public datasets
with varying target domains.
Empirical evaluation shows that 
in comparison to shallow fusion,
the proposed method
can provide a relative word error rate reduction (WERR)
by up to 9.8\% on unseen target domains.
%
%
\item \textbf{Zero-shot domain adaptation.}
Departing from conventional wisdom  
that ILME requires a \textit{target} domain LM 
trained on target domain data for inference,
we show that the proposed ILME technique
can provide up to 9.3\% relative WERR over shallow fusion
with only the \textit{source} domain LM,
thus not requiring 
any additional target domain data.
\item \textbf{ILME improves OOV F1.}
We show that the proposed approach
not only improves word error rate (WER),
but can also 
consistently improve the F1 score
for detecting out-of-vocabulary terms
from target domains
by up to 24.6\% relative to shallow fusion.
\end{itemize}

\section{Background} \label{sec:background}
\vspace{-0.5em}

Given an input sequence of speech features
$\mathbf{X}=\{\mathbf{x}_1, \ldots, \mathbf{x}_T\}$
and the corresponding output token sequence
$\mathbf{Y}=\{y_1, \ldots, y_L\}$,
the objective of an ASR model, $\theta$,
is to estimate the posterior distribution
$P(\mathbf{Y} | \mathbf{X};\theta)$.
Typically, 
log-posteriors are computed
for numeric stability.
Finally, the optimal transcription $\hat{\mathbf{Y}}$
is determined from these log-posteriors
using an autoregressive beam search algorithm
with the criteria:
$
\hat{\mathbf{Y}} 
=
\argmax_{\mathbf{Y}}
\Big[ 
    \log P(\mathbf{Y}|\mathbf{X};\theta^\text{S}_\text{ASR}) 
\Big]
$.

\smallskip
\noindent \textbf{LM Fusion for Domain Adaptation}

\noindent
Also referred to as \textit{text-only} adaptation, 
in LM fusion, a separately trained target domain LM, ${\theta^\text{T}_\text{LM}}$,
is combined with the source domain ASR model, $\theta^\text{S}_\text{ASR}$,
to bias the system to output text with target domain semantics.
Shallow Fusion~\cite{hannun2014deep} is one of the most popular methods of LM fusion~\cite{le2021deep}.
In this approach, the log-posteriors of the source domain ASR model 
are interpolated with target domain LM log-probabilities 
during inference. 
The weighted log-scores are then used 
to select the output token at each step of beam search 
using the criteria:
$
\hat{\mathbf{Y}} 
=
\argmax_{\mathbf{Y}}
\Big[ 
    \log P(\mathbf{Y}|\mathbf{X};\theta^\text{S}_\text{ASR})
    + \lambda_T \log P(\mathbf{Y};\theta^\text{T}_\text{LM})
\Big]
$,
where 
$P(\mathbf{Y};\theta^\text{T}_\text{LM})$ 
is the probability assigned to the token sequence $\mathbf{Y}$ 
by the target domain LM, 
and $\lambda_T$ is the target LM weight.
In this work,
we compare the proposed ILME approach
with this widely used shallow fusion technique
for domain adaptation.


\smallskip
\noindent \textbf{ILM Estimation for AED and RNN-T Models}

\noindent
Internal LM estimation
has been proposed to reduce
implicitly learned model bias from
the source domain~\cite{meng2021internal}.
To estimate the internal LM,
joint softmax approximation~\cite{variani2020hybrid}
is invoked
to decompose the joint model parameters $\theta^S_{ASR}$
into individual AM and LM components.
Assuming domain invariant acoustic conditions,
a reformulation~\cite{meng2021internal} 
of the Bayes' theorem
for the posterior yields
the inference criteria:
\vspace{-0.8em}
\begin{align}
\hat{\mathbf{Y}} 
= 
\argmax_{\mathbf{Y}}
\Big[ 
    \log P(\mathbf{Y}|\mathbf{X};\theta^\text{S}_\text{ASR})
    &+ \lambda_T \log P(\mathbf{Y};\theta^\text{T}_\text{LM}) \label{eq:ilm-inference}  \\
    &- \lambda_I \log P(\mathbf{Y};\theta^\text{S}_\text{ASR}) \nonumber
\Big],
\end{align}
where $\lambda_I$ is internal LM weight,
and $\log P(\mathbf{Y};\theta^\text{S}_\text{ASR})$
is computed in an autoregressive manner
by completely eliminating the acoustic input.
This approach is limited to AED and RNN-T models
as they inherently perform autoregressive decoding, 
and it is straightforward to decouple
the model parameters
into AM and LM components.
In this work,
we propose a novel technique
for adapting the ILME algorithm
to the non-autoregressive decoding
of CTC models.


\vspace{-0.5em}
\section{Estimating ILM for CTC-based ASR} \label{sec:methodology}
\vspace{-0.5em}

The core principle of the original ILME technique
is to estimate the model's internal LM,
$P(\mathbf{Y}; \theta^\text{S}_\text{ASR})$,
in an autoregressive manner:
$
P(\mathbf{Y}; \theta^\text{S}_\text{ASR}) = \prod^{L}_{l=1}P(y_l | \mathbf{Y}_{1:l-1}; \theta^\text{S}_\text{ASR})
$,
which lends itself naturally 
to AED and RNN-T models.
As these models perform autoregressive decoding,
they consist of a parametric component
that can be identified as learning probabilities over token sequences
--- 
for AED models this would be the decoder network,
and for RNN-T models it would be the prediction network.
In contrast, CTC models compute log-posteriors
in a non-autoregressive fashion,
i.e.,
we have $\mathbf{Y}=\{y_1, \ldots, y_T\}$ output tokens
(including \textit{blank} tokens)
corresponding to each input timestep $\mathbf{x}_t$.
Hence, there is no separate parametric component
for tokens.
The CTC loss also has an inherent conditional independence
between the output tokens $\mathbf{Y}$,
given the input $\mathbf{X}$.
However, the self-attention and convolution mechanisms
in transformer and conformer-based neural architectures
implicitly relaxes the conditional independence,
due to the sequence-level interdependence of the intermediate features
from attention and convolution.
Therefore, CTC models still inadvertently
learn an internal LM.
An example of this is shown
in \cref{tab:ilm-example}.
Additionally,
the non-autoregressive architecture of CTC models
makes it inconceivable to structurally decompose
the model into corresponding acoustic and language components,
which is a precursor to eliminating the acoustic input
for conventionally estimating the internal LM.

Fundamentally,
the ILM is the model's bias
in the absence of any acoustic context.
Consequently, 
we overcome the above challenges
for ILME with CTC models
by proposing an iterative input masking approach.
At a high level, we mask the input audio multiple times
and iteratively perform forward pass
to determine the CTC log-posteriors
for the corresponding masked timesteps.
We then accumulate the log-posteriors 
for different groups of masked timesteps
and normalize them in the posterior domain
to compute the final internal LM distribution.
We now explain our approach in further detail.

\smallskip
\noindent \textbf{Iterative Input Masking}

\noindent
Since it is not straightforward to decouple
the acoustic and language components of a CTC model,
we cannot simply zero-out the entire acoustic input
for estimating the ILM.
Hence, instead of completely eliminating the acoustic input,
we propose to iteratively mask the input audio
for performing ILME.
Given an input speech sequence 
$\mathbf{X}=\{\mathbf{x}_1, \ldots, \mathbf{x}_T\}$, 
we divide the input timesteps 
into $K$ equal, non-overlapping partitions,
and eliminate the input for each partition,
yielding a set of $K$ masked sequences
$\{\mathbf{\tilde{X}}_1, \ldots, \mathbf{\tilde{X}}_K\}$,
such that
\vspace{-0.5em}
\begin{align}
\mathbf{\tilde{x}}^{(k)}_t
&=
\begin{cases}
0, &\text{if }
t_{k-1} \leq t < t_k \\
\mathbf{x}_t & \text{otherwise}
\end{cases}
\end{align}

\noindent
where 
$\mathbf{\tilde{x}}^{(k)}_t$ is the input at $t$
for $\mathbf{\tilde{X}}_k$,
also $t_0 = 1$ and $t_k = kT/K$.

There are several masking strategies that can be followed here,
such as masking at word boundaries,
but experimentally we find that simply dividing the audio
into equal partitions yields best results.
Next, we discuss how we leverage these masked sequences
for ILME.


\smallskip
\noindent \textbf{Computing ILM for CTC}

\noindent
Given a vocabulary of $N$ tokens
$\{\mu_1, \ldots, \mu_N\}$,
we first denote the log-posteriors
computed by the CTC model
at timestep $t$ as:
\vspace{-0.5em}
\begin{align}
{\boldsymbol\Psi}_t(\mathbf{X}) 
&= 
\begin{bmatrix}
    \log P(y_t = \mu_1 |\mathbf{X};\theta^\text{S}_\text{CTC}) \\
    \vdots \\
    \log P(y_t = \mu_N|\mathbf{X};\theta^\text{S}_\text{CTC})
\end{bmatrix}
\end{align}

Next, we compute the log-posterior distribution vector
for a given masked sequence
${\boldsymbol\Psi}_t(\mathbf{\tilde{X}}_k)$.
Now,
we want to determine whether these log-probabilities
correspond to the original acoustic input
or if they are affected by the acoustic masking.
In the latter case, it would indicate the model's bias
when no acoustic information was passed,
and hence it would correspond to the model's ILM.
For this, we determine 
the element-wise absolute difference
between the log-posterior distributions as
${\boldsymbol\epsilon}^k_t = | {\boldsymbol\Psi}_t(\mathbf{\tilde{X}}_k) - {\boldsymbol\Psi}_t(\mathbf{X}) |$.
Intuitively, 
if all elements of the vector 
${\boldsymbol\epsilon}^k_t$
have relatively low values,
say below some threshold $\gamma$,
it means timestep $t$ is unaffected by the $K^\text{th}$ masking,
and we can ignore this in the ILM computation.
Conversely, we want to keep ${\boldsymbol\Psi}_t(\mathbf{\tilde{X}}_k)$
for the ILM computation
if ${\boldsymbol\epsilon}^k_t$
has any relatively high values,
as it indicates the model bias 
that is affected by the masking.

Hence, we take the maximum
of the $N$ values
in the vector ${\boldsymbol\epsilon}^k_t$,
denote it as $\delta^k_t$,
and normalize it across all timesteps 
to determine a scalar $\hat{\delta}^k_t$
that we can compare against a threshold $\gamma$
for determining whether 
${\boldsymbol\Psi}_t(\mathbf{\tilde{X}}_k)$
will contribute to the ILM.
More concretely:
\vspace{-0.5em}
\begin{align}
\delta^k_t = \max_n \big( {\boldsymbol\epsilon}^k_t [n] \big);
\quad
\hat{\delta}^k_t &= \frac{\delta^k_t}{\max_{t'}(\delta^k_{t'})}
\end{align}

Finally, 
we estimate the ILM for timestep $t$
by computing a pseudo log-likelihood
using the log-posterior distributions
across all $K$ masks 
conditioned on whether 
the corresponding $\hat{\delta}^k_t$ 
is above a threshold $\gamma$.
Therefore,
\vspace{-1em}
\begin{align}
\log P_\text{ILM} (\mathbf{y}_t;\theta^\text{S}_\text{CTC}) &= \text{LogSoftmax} \Bigg( \sum_{k=1}^K {\boldsymbol\Psi}_t(\mathbf{\tilde{X}}_k) \Big[\hat{\delta}^k_t > \gamma\Big] \Bigg) \label{eq:ilm-pseudo-ll}
\end{align}

Although we show the ILM computation for a single timestep $t$,
all the steps discussed are easily vectorizable
in the sequence dimension,
and can be computed for all timesteps in parallel
using efficient sub-linear methods.
The main overhead of our approach
comes from the multiple forward passes
required to compute
the log-posteriors
across all masks,
which introduces a cost
that scales linearly with $K$.
This can be mitigated in part
by doing one batched forward pass
with the original and masked inputs
in the same batch.


\smallskip
\noindent \textbf{Inference with ILM Pseudo Log-Likelihood}

\noindent
For inference with ILME,
we follow a similar approach as 
described in \cref{eq:ilm-inference},
subtracting the ILM pseudo log-likelihood
computed using \cref{eq:ilm-pseudo-ll}
from the original CTC log-posterior.
However,
CTC models have a special \textit{blank} token
that is purely functional 
and does not correlate
with any language semantics.
Hence, similar to \cite{meng2021internal},
we skip subtracting the ILM
for the timesteps
where blank token is originally predicted.
However, given the peaky behaviour of CTC models~\cite{zeyer2021does},
we skip the timesteps for blank token
only when the model assigns it a high likelihood
in the posterior domain,
above some threshold $\beta$:
\vspace{-0.55em}
\begin{align}
\text{if} \
&P_\text{CTC}(y_t = \langle\texttt{blank}\rangle|\mathbf{X}) < \beta, \nonumber \\
&\textsc{Score}(\mathbf{y_t})
=
\log P_\text{CTC}(\mathbf{y}_t|\mathbf{X}) 
- \lambda_I \log P_\text{ILM} (\mathbf{y}_t) \\
\text{else,} \
&\textsc{Score}(\mathbf{y_t}) = \log P_\text{CTC}(\mathbf{y}_t|\mathbf{X})
\end{align}
Finally, we pass these modified log-posterior scores, $\textsc{Score}(\mathbf{y_t})$,
to beam search for performing inference combined with shallow fusion.


\vspace{-0.5em}
\section{Experiments} \label{sec:experiments}
\vspace{-0.5em}

\noindent \textbf{Model.}
We perform all our experiments on a conformer-based model~\cite{Gulati2020},
 trained in a hybrid fashion~\cite{watanabe2017hybrid}
using joint CTC and decoder attention loss for more stable training.
During inference, 
we only use the non-autoregressive CTC head
and discard the shallow decoder.
The conformer encoder consists of 
$20$ layers of conformer blocks, 
where each layer has $8$-headed attention 
and $2048$ hidden units 
followed by a $512$-dimensional output projection. 
The shallow decoder is a single transformer-based layer
with a hidden dimension of $2048$.
The model has approximately 140M parameters, 
and is trained using the ADAM optimizer
with an initial learning rate of $3\times10^{-3}$ for 45 epochs.
For tokenization, we train a sentencepiece tokenizer 
with a vocabulary size of $2048$.
Our model implementation and training is done by leveraging
the widely used ESPnet framework~\cite{Watanabe2018}.
In addition, we train 4-gram language models (LMs) 
for shallow fusion experiments and ILME
using a modified Kneser-Ney smoothing algorithm
implemented in the popular KenLM library~\cite{heafield2011kenlm}. 

\smallskip
\noindent \textbf{Data.}
The model is trained 
on a large English corpus of 50k+ hours paired audio text data. 
We sample this data from public and in-house paired audio and text, 
ensuring a good mix of accents, speakers, sampling rates and background noise. 
The data regime is representative of a wide range of end-to-end ASR systems for various speech applications.
To evaluate the proposed ILME approach, 
we consider four evaluation datasets from varying domains: 
LibriSpeech, VoxPopuli, Wikipedia and WSJ. 
We use the official test splits 
for each datasets in our inference experiments.
\textbf{LibriSpeech}~\cite{Panayotov2015} 
is a read English speech corpus based on LibriVox audiobooks. 
We consider the two official evaluation sets: 
\textit{test-clean}
and \textit{test-other},
each with 5 hours of test audio. 
\textbf{VoxPopuli}~\cite{Wang2021} 
consists of public political speech, 
sampled from 2009-2020 European Parliament event recordings. 
For our evaluation purpose, we utilize a 5-hour subset of VoxPopuli English data. 
The \textbf{Wikipedia} evaluation set~\cite{das2022listen} 
is a custom text-to-speech (TTS) dataset, 
which consists of a wide range of spoken named entities.
More specifically, 14k sentences are sampled from English Wikipedia
and passed through a TTS service
to produce synthetic audios of 33 hours with different speakers.
The \textbf{Wall Street Journal (WSJ)}~\cite{LDC93S6A} corpus 
contains conventional and spontaneous dictation by journalists.
The \textit{test\_eval92} split of 0.7 hours is selected for our evaluation.

Our AM training data has no overlap whatsoever with 
data from the evaluation sets.
Hence all 4 evaluation sets are regarded as \textit{out-of-domain}.
\cref{tab:test-sets} shows a summary of these evaluation sets.
Also, when we train a \textit{target} LM, 
only the corresponding training splits 
of LibriSpeech, VoxPopuli, Wikipedia and WSJ are used respectively.

\begin{table}[t!]
\footnotesize
\centering
\caption{Summary of out-of-domain evaluation datasets.}
\vspace{-1em}

\begin{tabular}{@{}lcccc@{}}
\toprule
Dataset                       & Domain        & \# Hrs.   & \# OOVs   & \# Unique \\
\midrule
LibriSpeech (test-clean)      & Audiobooks    & 5.4     & 232         & 8138     \\
LibriSpeech (test-other)      & Audiobooks    & 5.3     & 265         & 7597     \\
VoxPopuli                     & Parliamentary & 4.9     & 52          & 5168     \\
Wikipedia                     & Encyclopedia  & 32.7    & 9233        & 38706    \\
WSJ                           & News          & 0.7     & 0           & 1834     \\
\bottomrule
\end{tabular}
\label{tab:test-sets}
\vspace{-2em}
\end{table}

\smallskip
\noindent \textbf{Experimental Setup.}
In order to demonstrate the generalizability of the proposed approach,
we perform all experiments with a single set of hyperparameters
across all datasets,
exhaustively tuned on an in-house development set.
For all experiments,
we perform CTC inference using the beam search algorithm
with a beam size of 50.
For LM fusion,
we use an LM weight of $\lambda_T=1.0$.
While performing ILME, we mask the audios into $K=5$ partitions
and perform inference with an ILM weight of $\lambda_I=0.1$,
a log-posterior delta threshold of $\gamma=0.25$
and a blank-token filter threshold of $\beta=0.9$.

\vspace{-1.8em}
\section{Results} \label{sec:results}
\vspace{-0.5em}

We showcase the efficacy 
of the proposed ILME approach
on multiple out-of-domain datasets
with a baseline of CTC Beam Search (BS)
as well as
with stronger Shallow Fusion (SF) inference
with an external LM.
We perform our multifaceted evaluation in two modes:
\begin{enumerate}[leftmargin=*,topsep=1pt,itemsep=0pt]
\item \textbf{text-only domain adaptation},
in which we use a \textit{target} LM trained on target domain text.
This is typical in studying the efficacy of ILME domain adaptation
in previous works~\cite{meng2021internal}.
\item \textbf{zero-shot domain adaptation},
in which we only use a \textit{source} LM trained on source domain text,
i.e., only the audio transcriptions already available for training the ASR model.
\end{enumerate}
Note that the latter case is a much stronger
test of generalization
as the inference does not include any information
about the target domain.
To further calibrate generalization performance, 
we also mine out-of-vocabulary (OOV) terms 
from each out-of-domain evaluation dataset,
and evaluate the F1 score
for detecting these OOV terms
which were never encountered by the model during training.
\cref{tab:test-sets} also shows 
the number of OOV references
found in the evaluation datasets.


\begin{table*}[t!]
\footnotesize
\centering
\caption{
WER (lower is better) and relative WERR (shown in parenthesis). 
}
\vspace{-1em}
\begin{tabular}{@{}ccc|r@{\hspace{0.4\tabcolsep}}r|r@{\hspace{0.4\tabcolsep}}r|r@{\hspace{0.4\tabcolsep}}r|r@{\hspace{0.4\tabcolsep}}r|r@{\hspace{0.4\tabcolsep}}r@{}}
\toprule
                                                                                                   & Method  & LM         & \multicolumn{2}{c|}{\begin{tabular}[c]{@{}c@{}}LibriSpeech\\ test-clean\end{tabular}}  & \multicolumn{2}{c|}{\begin{tabular}[c]{@{}c@{}}LibriSpeech\\ test-other\end{tabular}}  & \multicolumn{2}{c|}{VoxPopuli}         & \multicolumn{2}{c|}{Wikipedia}         & \multicolumn{2}{c}{WSJ}              \\
\midrule
                                                                                                   & BS      & no LM      & 4.0                                     & \multicolumn{1}{c|}{-}                       & 8.1                                     & \multicolumn{1}{c|}{-}                       & 9.9          & \multicolumn{1}{c|}{-}  & 13.0         & \multicolumn{1}{c|}{-}  & 3.4          & \multicolumn{1}{c}{-} \\
\midrule
\multirow{2}{*}{ \begin{tabular}[c]{@{}c@{}}Zero-shot\\ Domain Adaptation\end{tabular}} & SF      & source LM  & 4.1                                     & (-2.5\%)                                     & 7.9                                     & (+2.5\%)                                     & 9.3          & (+6.1\%)                & 10.8         & (+16.9\%)               & 3.6          & (-5.9\%)              \\
                                                                                                   & ILME    & source LM  & 3.7                                     & (+7.5\%)                                     & 7.6                                     & (+6.2\%)                                     & 9.3          & (+6.1\%)                & 10.7         & (+17.7\%)               & 3.0          & (+11.8\%)             \\
\midrule
\multirow{2}{*}{ \begin{tabular}[c]{@{}c@{}}Text-only\\ Domain Adaptation\end{tabular}} & SF      & target LM  & 3.2                                     & (+20.0\%)                                    & 6.6                                     & (+18.5\%)                                    & 7.9          & (+20.2\%)               & 9.5          & (+26.9\%)               & 3.2          & (+5.9\%)              \\
                                                                                                   & ILME    & target LM  & \textbf{2.9}                            & \textbf{(+27.5\%)}                           & \textbf{6.5}                            & \textbf{(+19.8\%)}                           & \textbf{7.9} & \textbf{(+20.2\%)}      & \textbf{9.3} & \textbf{(+28.5\%)}      & \textbf{2.9} & \textbf{(+14.7\%)}    \\
\bottomrule
\end{tabular}
\label{tab:wer}
\vspace{-1.5em}
\end{table*}
\begin{table}[t!]
\footnotesize
\centering
\caption{F1 (higher is better) for detecting OOVs using \textit{target} LM.}
\vspace{-0.5em}
\begin{tabular}{lccc}
\toprule
Dataset                  & BS   & SF   & ILME          \\
\midrule
LibriSpeech (test-clean) & 30.7 & 32.5 & \textbf{43.1} \\
LibriSpeech (test-other) & 33.3 & 27.4  & \textbf{33.3} \\
VoxPopuli                & 31.7 & 34.4 & \textbf{44.1} \\
Wikipedia                & 39.4 & 42.4 & \textbf{47.6} \\
\bottomrule
\end{tabular}
\label{tab:f1-tgt}
\vspace{-0.3em}
\end{table}

\vspace{-1em}
\subsection{Text-only Domain Adaptation}
\vspace{-0.5em}

\noindent \textbf{Transcription WER.}
In \cref{tab:wer},
we show the WER and relative WERR
for each inference technique.
For text-only domain adaptation
using the \textit{target} LM,
we clearly see that the proposed ILME approach
outperforms the baseline in every case,
showing a relative WERR by up to 28.5\% (Wikipedia).
ILME also outperforms the shallow fusion approach
for the LibriSpeech~test-\{clean, other\}, Wikipedia and WSJ datasets,
improving the WER relative to SF by up to 9.4\% (LibriSpeech~test-clean and WSJ).
For VoxPopuli, even though WER improvement 
for ILME is very similar to SF,
we can see a clear benefit of using ILME
in terms of OOV F1 score as seen in \cref{tab:f1-tgt}.

\smallskip
\noindent \textbf{F1 score for detecting OOVs.}
\cref{tab:f1-tgt} shows the F1 score
for detecting OOV terms using the \textit{target} LM 
for the out-of-domain datasets.
As there are no OOV terms found in WSJ,
we exclude that dataset from this evaluation.
We see that ILME outperforms both BS and SF
for multiple datasets: 
LibriSpeech~test-clean, VoxPopuli and Wikipedia.
ILME improves over the baseline method's F1 score
relatively by up to 40.4\% (LibriSpeech~test-clean).
For VoxPopuli, we see a 39.1\% relative improvement in F1 from BS,
whereas SF only improves F1 relatively by 8.5\% from BS.
This shows ILME is clearly a better choice than SF for VoxPopuli
even though they have very similar WER.
In the case of LibriSpeech~test-other,
we see that ILME performs the same as BS.
However, looking at the shallow fusion F1 score 
for LibriSpeech~test-other,
it is evident that the target LM actually degrades
the OOV detection performance.
This can be attributed to the fact
that LibriSpeech~test-other 
is deliberately more challenging~\cite{Panayotov2015},
and we note that the LibriSpeech training data
has extremely sparse representation of the OOV terms.
Hence, it is not guaranteed 
that using a target LM by itself
can improve the detection of OOV terms,
as these rare terms 
are not guaranteed to be present
in the target domain text.
However, 
ILME is still able to
recover the degradation 
in shallow fusion F1 
using the target LM,
while also improving WER
for LibriSpeech~test-other.
ILME is able to this
by removing the model bias at each timestep,
that would allow the OOV tokens to surface up
compared to the previously biased tokens.


\vspace{-1em}
\subsection{Zero-shot Domain Adaptation}
\vspace{-0.5em}

\begin{table}[t!]
\footnotesize
\centering
\caption{F1 (higher is better) for detecting OOVs using \textit{source} LM.}
\vspace{-0.5em}
\begin{tabular}{lccc}
\toprule
Dataset                  & BS   & SF   & ILME          \\
\midrule
LibriSpeech (test-clean) & 30.7 & 36.6 &  \textbf{45.8} \\
LibriSpeech (test-other) & 33.3 & 32.8 & \textbf{39.4} \\
VoxPopuli                & 31.7 & 26.2 & \textbf{34.4} \\
Wikipedia                & 39.4 & 39.5 & \textbf{45.2} \\
\bottomrule
\end{tabular}
\label{tab:f1-src}
\vspace{-1.5em}
\end{table}

In real world ASR applications, 
it is often hard to determine 
the specific domain for the speech input apriori. 
This prompted us to evaluate the generalization of ILME 
in situations where we are not aware 
of the target domain, 
or we have no data from the target domain to train an LM.
For this, we use a \textit{source} LM
trained on the ASR training transcriptions.
In using the source LM directly without ILME, 
we run the risk of doubly biasing the output
due to the internal model bias 
combined with the external LM bias.
Thus, ILME aids zero-shot domain adaptation 
by mitigating the internal bias of the source domain,
while the external source LM can ensure 
that higher level semantics
pertaining to the language itself
are preserved.

\smallskip
\noindent \textbf{Transcription WER.}
\cref{tab:wer} shows that
in the case of zero-shot domain adaptation
using a \textit{source} LM,
ILME outperforms BS in all cases
with up to 17.7\% (Wikipedia) relative WERR.
As we are using a \textit{source} LM in this case,
simple shallow fusion may not yield better results than BS,
as can be seen in the case of 
LibriSpeech~test-clean (-2.5\%) and WSJ (-5.9\%).
This is because the inference may get doubly biased
from the source domain data by the
internal as well as external LM.
ILME can help in such cases
by removing the internal bias of the source domain.
This can be noticeably seen in \cref{tab:wer},
as ILME outperforms SF using only the \textit{source} LM
for multiple datasets: 
LibriSpeech~test-\{clean, other\}, Wikipedia and WSJ. 
For VoxPopuli, again, ILME performs similar to SF,
but ILME is once more clearly the optimal choice 
when the OOV F1 score is considered (\cref{tab:f1-src}).

\smallskip
\noindent \textbf{F1 score for detecting OOVs.}
In \cref{tab:f1-src}, we show the F1 score
for detecting OOV terms using the \textit{source} LM.
We see that ILME unambiguously outperforms
the other techniques in all the cases of this
zero-shot domain adaptation scenario,
improving F1
relatively by up to 49.2\% (LibriSpeech~test-clean).
For the LibriSpeech datasets,
we see that ILME as well as SF
with the \textit{source} LM
is actually better at detecting OOV terms
than the \textit{target} LM.
This is because
there is no guarantee that the OOV terms
would be present in the training data
for the target LM,
as LibriSpeech samples the training data
from a separate partition of speakers and audiobooks~\cite{Panayotov2015}.
For VoxPopuli, again,
while ILME performs similar to SF
in terms of WER,
ILME is clearly the better option
as it improves F1 relatively by 8.5\%
whereas SF actually degrades 
F1 by 17.4\%.
Coincidentally for VoxPopuli,
ILME with \textit{source} LM
performs as well as SF with \textit{target} LM.

\vspace{-1em}
\subsection{Future Work} \label{sec:discussion}
\vspace{-0.5em}

We observe that
previous ILME methods for AED and RNN-T models
only perform decoder-side de-biasing by eliminating the encoder input.
In contrast, our CTC-based ILME method works for
non-autoregressive decoding
using the encoder itself.
Hence, we hypothesize that our ILME approach can be extended
to AED and RNN-T models as well
for complementary encoder-side de-biasing.
Furthermore,
post-training techniques
can be formulated
for training student models
to estimate the ILM of a CTC model,
thus accelerating ILME inference.
We aim to pursue these directions in future work.

\vspace{-0.5em}
\section{Conclusion} \label{sec:conclusion}
\vspace{-0.5em}

In this work,
we propose a novel ILME approach
for CTC-based ASR models.
We extensively evaluate the efficacy 
of the approach
in adapting to multiple unseen target domains.
We show that the proposed ILME technique
improves 
WER by up to 9.8\%
and OOV F1-score
by up to 24.6\% relative to shallow fusion,
when only target domain text data is available.
We also demonstrate that applying ILME is beneficial
in application to any new target domain
even when there is no access to target domain data at all,
as it de-biases the encoder and provides out of the box improvement.

\vfill\pagebreak

\bibliographystyle{IEEEtran}
\bibliography{refs}

\end{document}